\documentclass[onecolumn,floatfix,superscriptaddress,a4paper,
               showpacs,showkeys,nofootinbib,preprint]{revtex4}
\usepackage{epsfig}%
\usepackage{latexsym}
\usepackage{xspace}
\usepackage{hyperref}
\usepackage[latin2]{inputenc}
\usepackage{indentfirst}
\usepackage{enumerate}

\usepackage{amsmath}
\usepackage{amssymb}
\usepackage[english]{babel}
\usepackage{url}

\newcommand{\eq}[1]{\begin{align} #1 \end{align}}

\begin{document}
\title{Statistical hadron-gas treatment of systems created \\ in
proton-proton interactions at CERN SPS}

\author{V. V. Begun}
\affiliation{
Warsaw University of Technology, Faculty of Physics, Koszykowa 75, 00-662 Warsaw, Poland}
\author{V. Vovchenko}
\affiliation{
Institut f\"ur Theoretische Physik, Goethe Universit\"at Frankfurt, D-60438 Frankfurt am Main, Germany}
\affiliation{
Frankfurt Institute for Advanced Studies, Giersch Science Center, D-60438 Frankfurt, Germany}
\author{M. I. Gorenstein}
\affiliation{Bogolyubov Institute for Theoretical Physics, 03680 Kiev, Ukraine}
\affiliation{
Frankfurt Institute for Advanced Studies, Giersch Science Center, D-60438 Frankfurt, Germany}
\author{H. Stoecker}
\affiliation{Institut f\"ur Theoretische Physik, Goethe Universit\"at Frankfurt, D-60438 Frankfurt am Main, Germany}
\affiliation{
Frankfurt Institute for Advanced Studies, Giersch Science Center, D-60438 Frankfurt, Germany}
\affiliation{
GSI Helmholtzzentrum f\"ur Schwerionenforschung GmbH, D-64291 Darmstadt, Germany}

\begin{abstract}
We analyze the newest data from the NA61/SHINE collaboration which, in addition to previous results on pions and kaons, include mean multiplicities of $p$, $\Lambda$, and  $\phi$-mesons produced in inelastic proton-proton (p+p) interactions at $\sqrt{s_{NN}}=6.3-17.3$~GeV.
The canonical ensemble formulation of the ideal hadron resonance gas (HRG) model is used with exact conservation of net baryon number $B=2$, electric charge $Q=2$, and strangeness $S=0$.
The chemical freeze-out parameters in p+p interactions are obtained and compared to those in central nucleus-nucleus
collisions.
%
%
Several features  of p+p interactions at the CERN SPS within a 
statistical model are studied:
1) the inclusion of the $\phi$-meson yields in thermal fits worsens significantly the fit quality;
2) the data show large event-by-event multiplicity fluctuations in inelastic p+p interactions which can not be explained by a single fireball described by a statistical model; 
3) the fits within the canonical ensemble formulation of HRG do not give any improvement over the fits within the grand canonical ensemble formulation, i.e., there are  no indications for existence of a single statistical system in p+p inelastic interactions in the considered energy range.

\end{abstract}



\maketitle

\section{Introduction}
Statistical hadron-resonance gas (HRG) models appeared to be rather successful in description of hadron multiplicities produced in high-energy nucleus-nucleus (A+A) collisions, see, e.g., Refs.~\cite{Cleymans:1992zc,Sollfrank:1993wn,Becattini:1997ii,BraunMunzinger:1999qy,Yen:1998pa,Becattini:2000jw}. The HRG in the canonical ensemble (CE) formulation is also capable to describe hadron multiplicities produced in different kinds of elementary reactions -- proton-proton (p+p), proton-antiproton, and electron-positron \cite{Becattini:1995if,Becattini:2000pr,Becattini:1997rv}, see also~\cite{Das:2016muc,Oeschler:2017bwk,Andronic:2016nof,Vislavicius:2016rwi,Andronic:2017pug} and references therein.
The chemical freeze-out temperature has been found from fitting mean hadron multiplicities in all these high energy processes.

The NA61/SHINE collaboration is performing the scan of the beam energy and system size at the Super Proton Synchrotron (SPS) of the European Organization for Nuclear Research (CERN)~\cite{Ga:2009,NA61facility,NA61-p+p-mult}. The NA61/SHINE delivers more and more data on mean multiplicities which can be added to the analysis within the HRG, and influence the obtained freeze-out parameters.

In our recent paper \cite{VBG-2016} the CE formulation of the HRG model was used to analyze the hadron yield  data \cite{NA61-p+p-mult} of the NA61/SHINE collaboration at the SPS energy region.
The following set of hadron species measured in inelastic p+p interactions was considered: $\pi^+$, $\pi^-$, $K^+$, $K^-$, $\overline{p}$. Application of the CE formulation of the HRG model is motivated by the fact that the above multiplicities were all measured in the $4\pi$ acceptance.
A good description of the data was obtained, $\chi^2/N_{\rm dof}\sim 1$, while the values of the extracted  chemical freeze-out temperature parameter was found to be close to the one in A+A collisions.

In the present paper we add the newly measured $4\pi$ multiplicities of $p$, $\Lambda$, and $\phi$ in p+p interactions at the center of mass energies of $\sqrt{s_{NN}}=7.7,~8.8,~12.3,~17.3$~GeV reported in Refs.~\cite{NA61-1,NA61-2,NA61-3}. Thus, new and most complete CE statistical model analysis for the full hadron yields in p+p inelastic reactions is done for the SPS energy region.
We find that the strongest effect on the quality of the fit appears when the $\phi$-meson yields are included in the fit.
Therefore, we present the results of our new fits in the two different ways: ``with $\phi$'', when all particles are included, and ``no $\phi$'', for the same set of hadrons, but without the $\phi$-meson yields.
We also discuss the event-by-event fluctuations of number of  hadrons  measured in inelastic p+p reactions and compare them with the CE results. Finally, the CE fits are compared to those obtained in the grand canonical ensemble (GCE) formulation of otherwise the same HRG model.

The global fit to multiplicities in CE or GCE ``with $\phi$'' does not reproduce the data well. 
This may lead to the conclusion that the special conditions required for the use of CE or GCE are not verified in the experimental situation discussed here.
An alternative scenario, namely that not one, but two or even more small statistical micro-subclusters are formed in inelastic p+p interactions, with masses and charges fluctuating from event to event, can not be excluded. Each of such micro-subclusters in any given event may then be describable by the micro-canonical ensemble (MCE) with exact conservation of the subcluster's energy and conserved charges.
These ensembles of MCE subclusters with parameters fluctuating event-by-event may then be represented as - and look like - a single statistical system, only if special conditions of CE or GCE are fulfilled.
Thus, even if these special conditions are not fulfilled, the concept of statistical hadronization with MCE of multiple subclusters per event may still hold.

The paper is organized as follows. The CE formulation of HRG is considered in Sec.~\ref{hrg}. The results for the fits of hadron yields in inelastic p+p interactions are presented in Sec.~\ref{results}. The summary in Sec.~\ref{sum} closes the paper.

\section{Hadron Resonance Gas Model}\label{hrg}
\subsection{The model formulation}\label{subsec:themodel}

An overview of the HRG model can be found, e.g., in Refs.~\cite{THERMUS,SHARE}. We just briefly remind that in the GCE formulation of the HRG model the conserved charges, such as net baryon number $B$, electric charge $Q$, and strangeness $S$, are conserved on average, but can differ from one microscopic state to another. The fitting parameters in GCE HRG are the temperature $T$, baryon chemical potential $\mu_B$\footnote{The chemical potentials $\mu_S$ and $\mu_Q$ correspond to the conservation of strangeness and electric charge, respectively. They are found from the conditions of zero average net strangeness and fixed average proton to neutron ratio in the colliding nuclei.}, the system volume $V$, and the strangeness saturation parameter $\gamma_S$~\cite{Letessier:1998sz,Letessier:2005qe,Rafelski:2015cxa}.

The differences between calculations within the CE and the GCE appear when the number of particles with corresponding conserved charge is smaller or of the order of unity~\cite{CE,CE1,Becattini:1995if,Becattini:2000pr,CE3,Becattini:1997rv,Begun:2005qd}.
The CE treatment of p+p collisions means that the conserved charges, $B=2$, $Q=2$, and $S=0$, should be fixed for each microscopic state of the hadron system. The fitting parameters in the CE are $T$, $V$, and $\gamma_S$.
We use the radius $R$ calculated from $V\equiv 4\pi R^3/3$ instead of volume for convenient comparison with the typical nuclear sizes.
 
The quantum statistics was not taken  into account in~\cite{VBG-2016}, but is applied in our present studies.
Introducing vector notations for the set of conserved charges $\vec Q=(B,Q,S)$ and auxiliary angles $\vec \phi= (\phi_1, \phi_2,\phi_3)$ one can present the CE partition function of the quantum HRG with three exactly conserved charges as the following~\cite{Keranen:2001pr,Hauer:2007ju,Becattini:2008tx}:
\begin{eqnarray}\label{BSQPartFunc_j}
\mathcal{Z}(\vec Q) = \frac{1}{(2\pi)^3}
\int \limits_{-\pi}^{\pi}
  d \phi_1
 \int \limits_{-\pi}^{\pi}
  d \phi_2
  \int \limits_{-\pi}^{\pi}
  d \phi_3~
  \exp\left[-i\,\vec Q \vec \phi ~
 +\sum_{j} z_j\right]~,
\end{eqnarray}
where $z_j$ is one-particle partition function,
\begin{eqnarray}\label{zj}
z_j=\frac{g_j\,V}{2\pi^2}\sum_{n=1}(\mp 1)^{n+1}\frac{T\,m_j^2}{n^2}	
 	       K_2\left(\frac{n\,m_j}{T}\right)\,\lambda_j^n\,(\gamma_S^{n_s^j+n_{\bar{s}}^j})^n
 \exp\left(i\,n\,\vec q_j\vec \phi\right)~,
 \end{eqnarray}
$g_j$ and $m_j$  are, respectively, the degeneracy factor and mass of $j$th particle species, $\vec q_j=(b_j,q_j,s_j)$ is the corresponding conserved charges of $j$th hadron, $n_s^j$ and $n_{\bar{s}}^j$ are the numbers of strange quarks and anti-quarks, respectively in the $j$th hadron. The quantity $\lambda_j$ is the auxiliary parameter which is set to unity in the final equations. It is used to obtain the moments of the multiplicity distribution of particles type $j$. The $K_2$ in Eq.~(\ref{zj}) denotes the modified Bessel function.
The summation $\sum_j$ includes stable hadrons and resonances.
The summation over $n$ in Eq.~(\ref{zj}) with $(\mp 1)^{n+1}$ takes into account the quantum statistics: the upper and lower sign corresponds, respectively, to Fermi-Dirac and Bose-Einstein statistics.
The mean CE multiplicity $\langle N_j\rangle$  is calculated by the derivative of $\ln~\mathcal{Z}$ over $\lambda_j$.
The first term $n=1$ in Eq.~(\ref{zj}) gives the Boltzmann approximation.
For the obtained values of the freeze-out parameters in both A+A and p+p collisions the effects of quantum statistics were checked to be small. They are of the size of the uncertainty for the determination of HRG parameters.
In the present paper the effects of Bose statistics are kept only for pions.
Note also that the pion number fluctuations are more sensitive to quantum statistics~\cite{Begun:2005ah}.

The thermal fits can be significantly affected by the excluded volume~\cite{Rischke:1991ke,Yen:1997rv,Yen:1998pa} or van der Waals~\cite{Vovchenko:2016rkn} interactions between hadrons, as illustrated in Refs.~\cite{Vovchenko:2015cbk,Vovchenko:2016ebv} for A+A collisions. These effects are not considered in the present study and our present results do not reflect the systematic uncertainties associated with hadronic interactions.

The HRG model fits are  done by minimizing the $\chi^2$ per number of degrees of freedom $N_{\rm dof}$:
 \eq{\label{xi}
 \frac{\chi^2}{N_{\rm dof}}
 ~=~\frac{1}{N_{\rm dof}}\sum_{i=1}^N\frac{\left(\langle N_i^{\rm exp}\rangle~-~\langle N_i^{\rm HRG}\rangle\right)^2}{\sigma_i^2}~,
}
where $\langle N_i^{\rm exp}\rangle$ is the experimentally measured hadron multiplicity, and $\langle N_i^{\rm HRG}\rangle$ is the hadron multiplicity calculated in the HRG. The $N_{\rm dof}$ is the number of the data points minus the number of fitting parameters.
The uncertainty $\sigma_i^2$ for each particle $i$ is calculated as the sum of the corresponding statistical and systematic uncertainties $\sigma_i^2=(\sigma_i^{syst})^2+(\sigma_i^{stat})^2$ .
All stable hadrons and well established resonances (those marked with *** and ****) that are listed by the Particle Data Group~\cite{pdg} are included. The Breit-Wigner shapes of resonances with finite constant widths are also taken into account. The list of particles runs up to $f_2(2340)$ for mesons and up to $N(2600)$ for (anti-)baryons. We do not include hadrons with charm and bottom degrees of freedom which have a negligible effect on the fit results. The $\sigma$ meson ($f_0(500)$) and the $\kappa$ meson ($K_0^*(800)$) are removed from the particle list, because of the reasons explained in Refs.~\cite{GomezNicola:2012uc,Venugopalan:1992hy,Broniowski:2015oha,Pelaez:2015qba}.
A mean multiplicity $\langle N_i\rangle$ of an $i$th particle specie is calculated in HRG as a sum of the primordial mean multiplicity $\langle N^{\rm prim}_i\rangle$ and resonance decay contributions,
\eq{\label{eq:Ntot}
\langle N_i\rangle~ =~
\langle N^{\rm prim}_i\rangle~ +~ \sum_R \langle n_i \rangle_R \, \langle N^{\rm prim}_R\rangle~,
}
where $\langle n_i \rangle_R$ is the average number of particles
of type $i$ resulting from decay of resonance $R$.

Calculations are performed in the framework of our publicly-available \texttt{Thermal-FIST} package~\cite{ThermalFIST}, which contains the quantum statistical CE implementation of the statistical model.

\section{CE HRG results for inelastic  p+p  interactions}
\label{results}

\subsection{Comparison with Data}

\begin{figure}[ht!]
\centering
\includegraphics[width=1.0\textwidth]{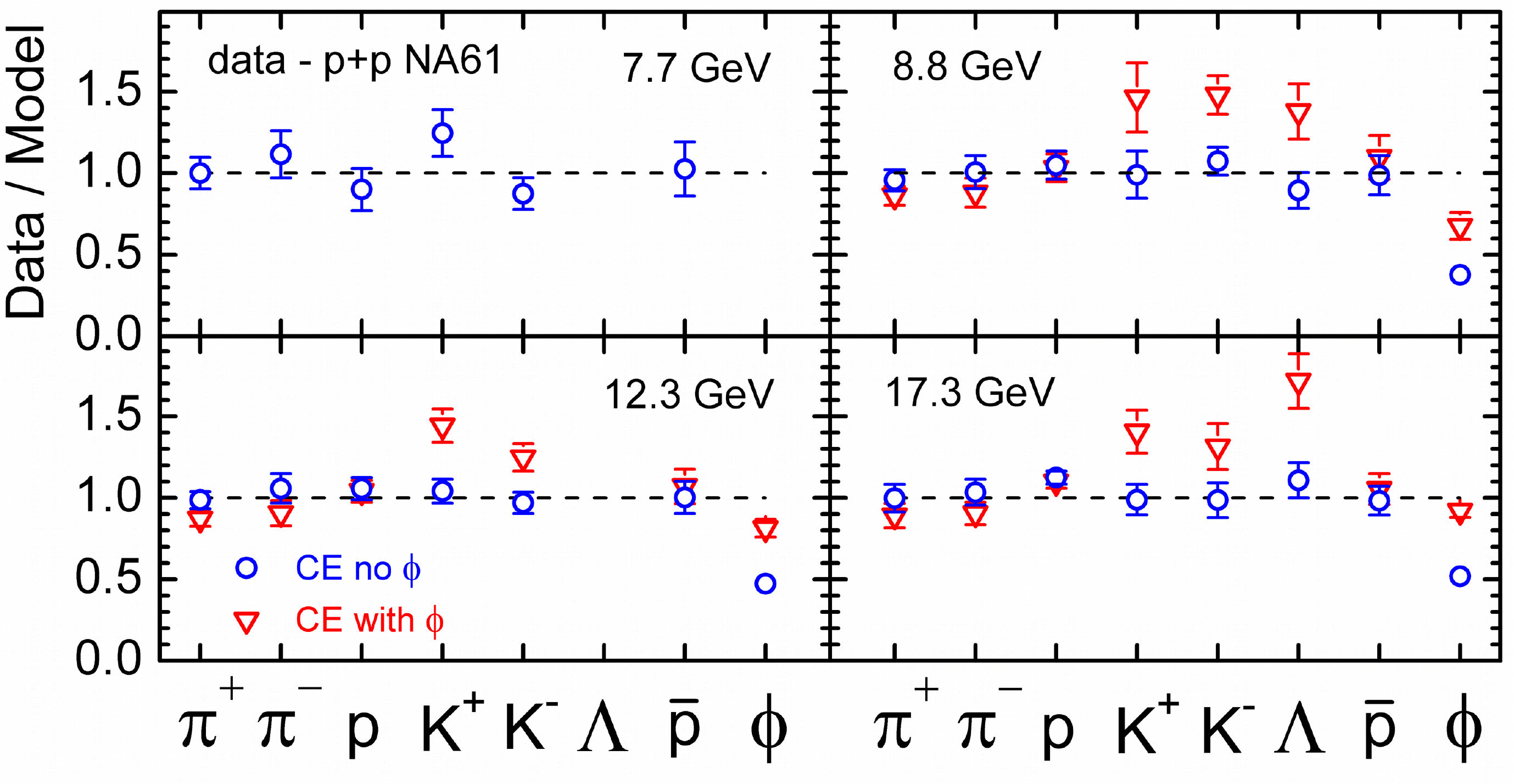}
\caption{The ratios of mean hadron multiplicities measured by the NA61/SHINE in p+p interactions at $\sqrt{s_{NN}}=7.7-17.3~$GeV (Data) to the mean multiplicities calculated from the best fit to the Data in the HRG model (Model).} \label{fig-NA61}
\end{figure}

\begin{figure}[ht!]
\centering
\includegraphics[width=0.49\textwidth]{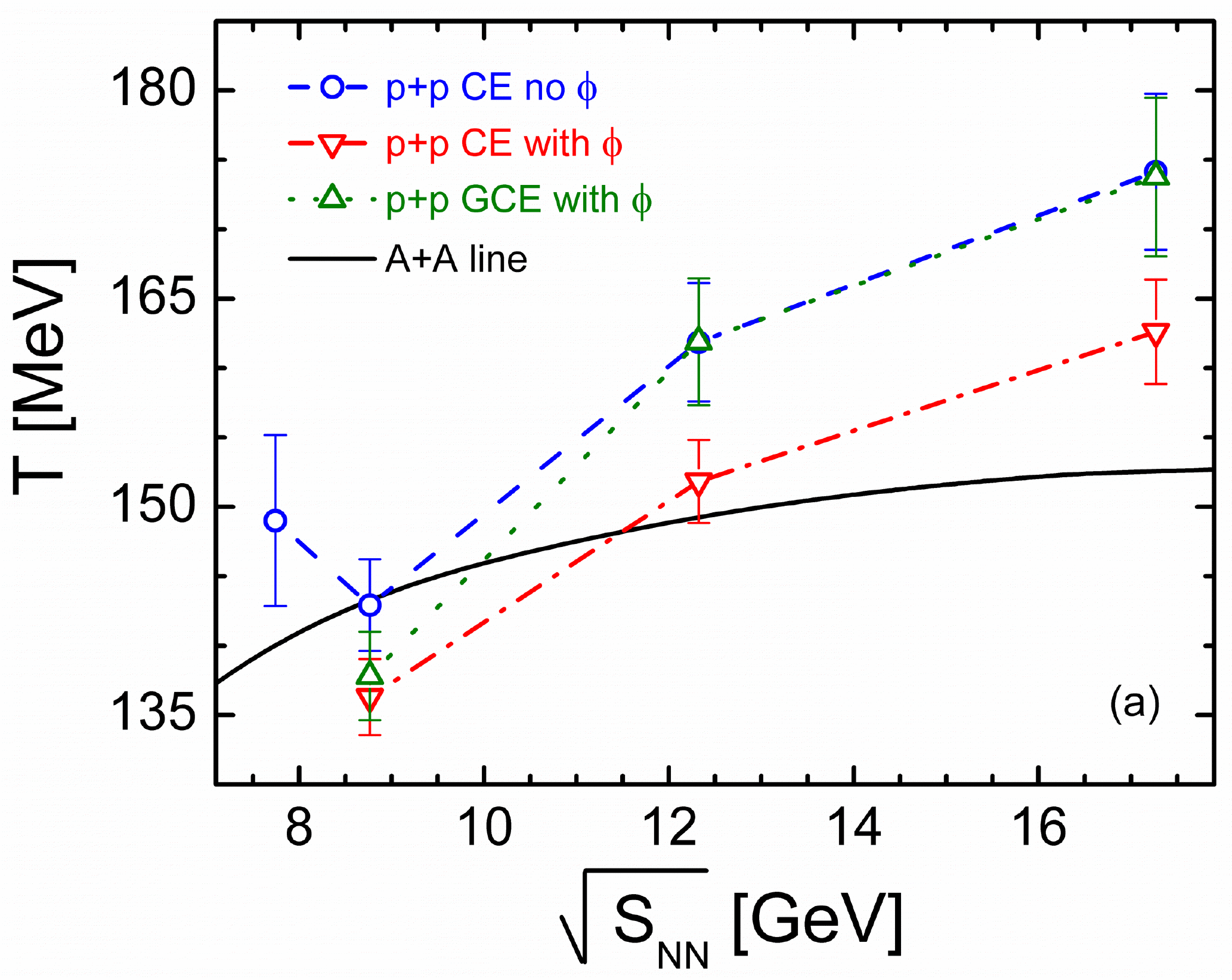}~~ 
\includegraphics[width=0.49\textwidth]{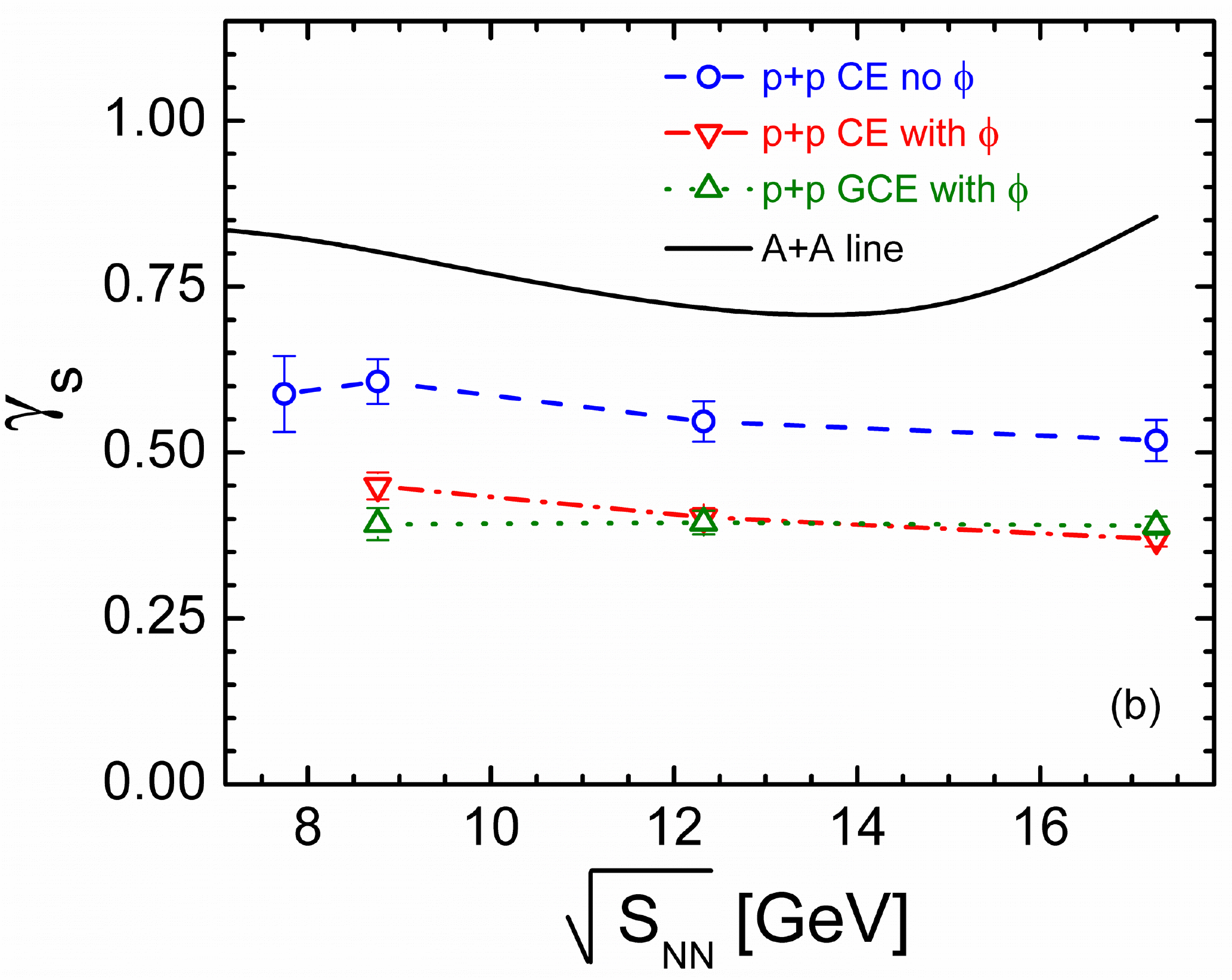} 
\\
\includegraphics[width=0.49\textwidth]{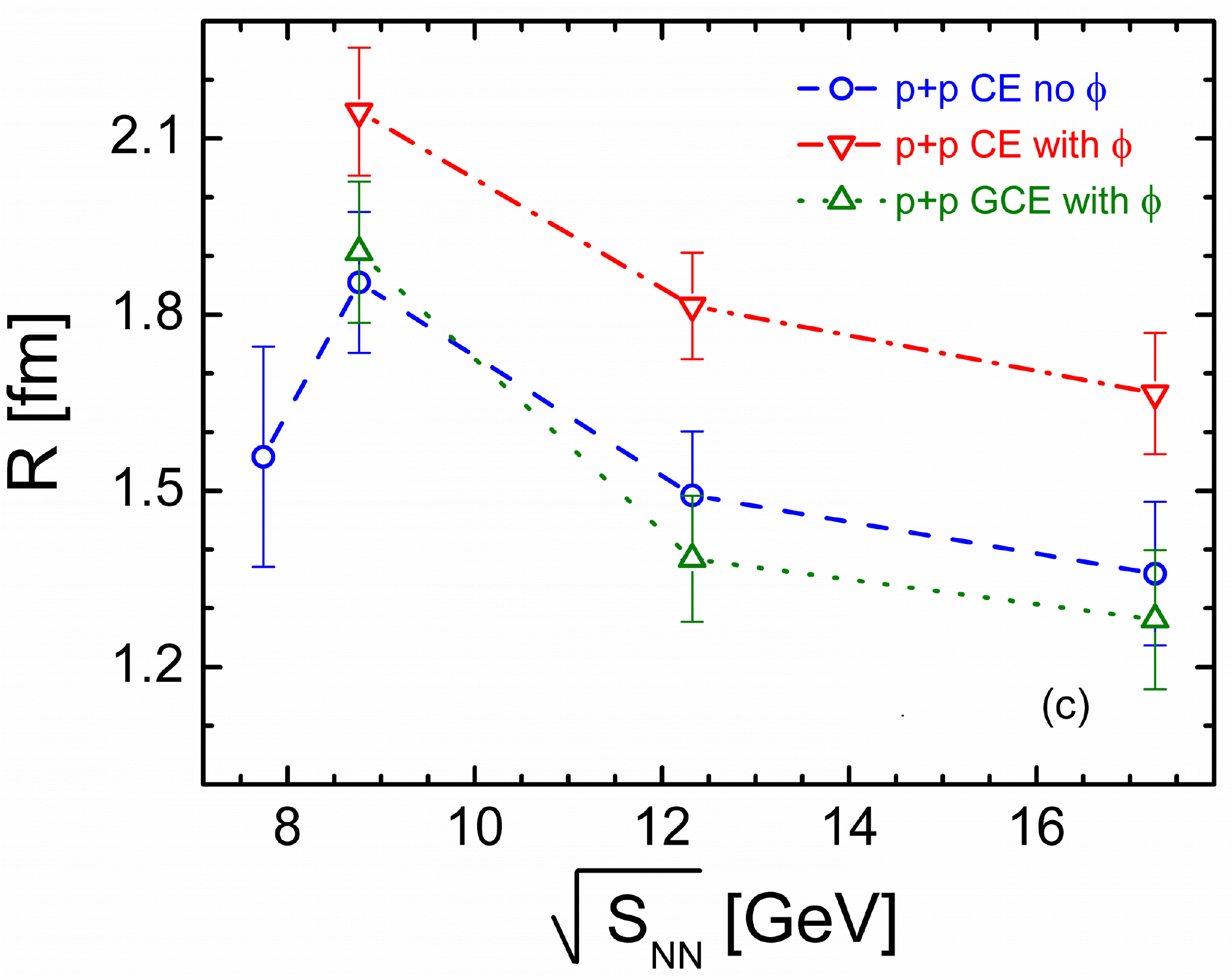}~~ 
\includegraphics[width=0.49\textwidth]{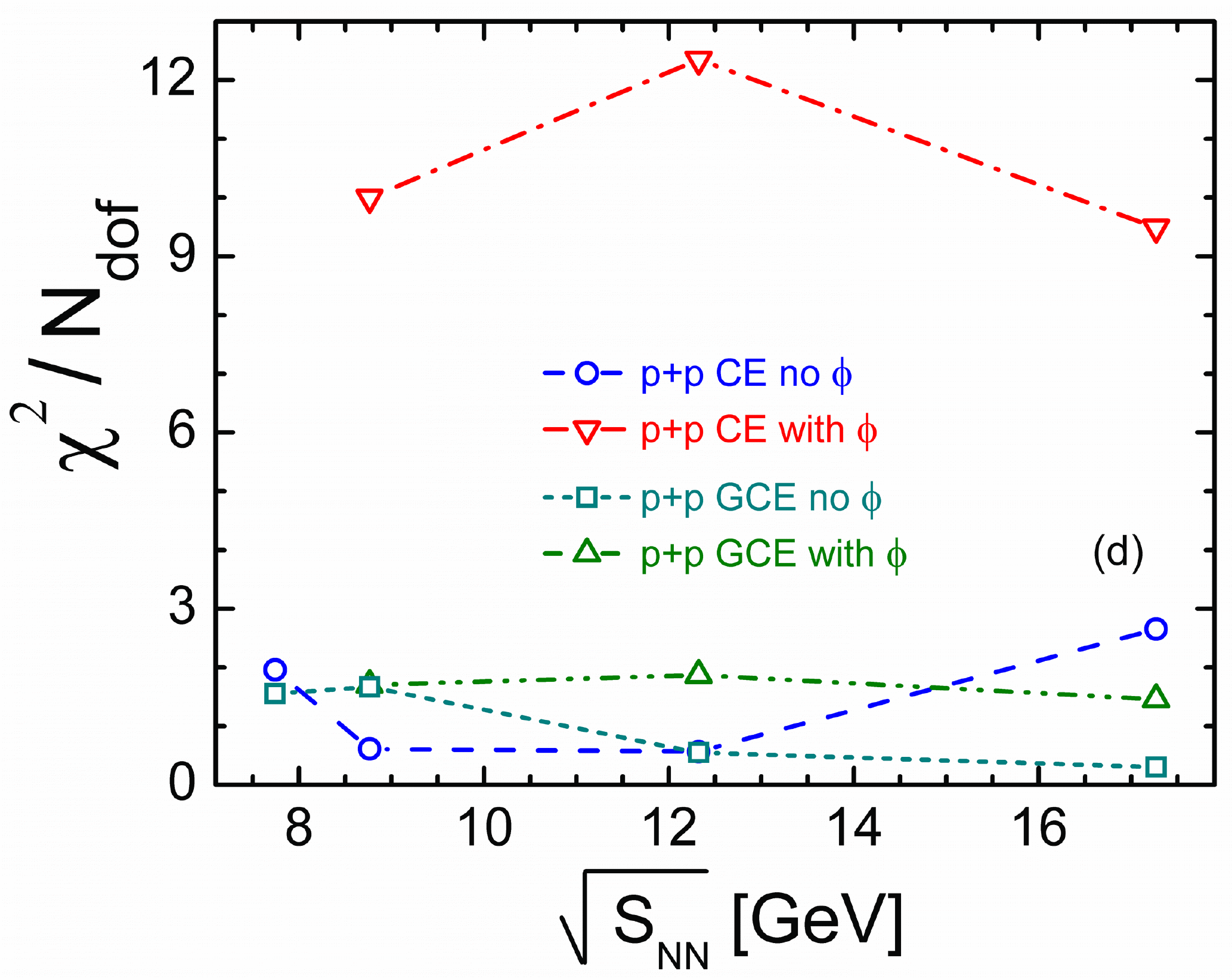} 
\caption{
Temperature $T$ (a),  strangeness saturation factor $\gamma_S$
(b), radius of the system $R$ (c), and  $\chi^2/N_{\rm dof}$ (d)  in inelastic p+p reactions
as functions of the collision energy. Solid lines in (a) and (b) show the corresponding
results in central Pb+Pb collisions from Ref.~\cite{VBG-2016}.} \label{fig-T}
\end{figure}

The hadron yield data of the NA61/SHINE collaboration for inelastic p+p interactions \cite{NA61-p+p-mult,NA61-1,NA61-2,NA61-3} are fitted within the CE HRG model. A comparison of the data with the HRG model fit is shown in Fig.~\ref{fig-NA61} and in Tables~\ref{Table:I} and  \ref{Table:II}. The obtained $T$, $\gamma_S$, and $R$ parameters are presented in  Fig.~\ref{fig-T} and Table~\ref{Table:III}. The solid lines in Figs.~\ref{fig-T} (a) and (b) depict the results for central Pb+Pb collisions from Ref.~\cite{VBG-2016}, and are shown for comparison. Other lines are made to guide the eye.
The missing points in Fig.~\ref{fig-NA61} mean that the corresponding particle multiplicity is not measured yet.
Inclusion of the $\phi$-meson yields to the fitted data leads to an improvement of the $\phi$ meson description, but it strongly worsens the description of other, mainly strange, particles. 

As seen in Fig.~\ref{fig-T} (a), the chemical freeze-out temperature $T$ in p+p interactions at different collision energies  remains close to the A+A values. A dependence of the $\gamma_S$ and $R$ parameters on collision energy in inelastic p+p interactions in the SPS energy region is seen in Figs.~\ref{fig-T} (b) and (d), respectively.
Note that the $R$ parameter should be interpreted with care, because  excluded volume corrections, which were neglected in the present paper, may strongly influence the total system volume~\cite{Begun:2012rf}: the values of $R$ would become larger, and their energy dependence could be changed.

\subsection{Statistical model approach to p+p data}
The behavior of the chemical freeze-out parameters presented in Figs.~\ref{fig-T} (a), (b), and (c) is rather similar to the one reported in Ref.~\cite{VBG-2016}. However, a more detailed analysis reveals several issues:
\\

1. The $\phi$ meson in inelastic p+p interactions.

Quality of the CE fits becomes much worse when the data on the $\phi$-meson multiplicity are included. This is clearly seen from Fig.~\ref{fig-T}~(d), where the results `with $\phi$' and `no $\phi$' (i.e., when the $\phi$ data are not included in the fit) can be compared. The values of $\chi^2/N_{\rm dof}$ significantly increase from $\chi^2/N_{\rm dof}\sim 1$ for `no $\phi$', to $\chi^2/N_{\rm dof}\sim 10$, when the $\phi$-meson yields are added to the fit. The experimental values for the $4\pi$-multiplicity of $\phi$ mesons in p+p collisions at SPS energies appear to be much smaller than the results of the CE fits.
This fact was first observed in Ref.~\cite{Bec} for the CE model analysis of the NA49 data in p+p collisions at $\sqrt{s_{NN}}=17.3$~GeV.

\begin{figure}[ht!]
\centering
\includegraphics[width=1.0\textwidth]{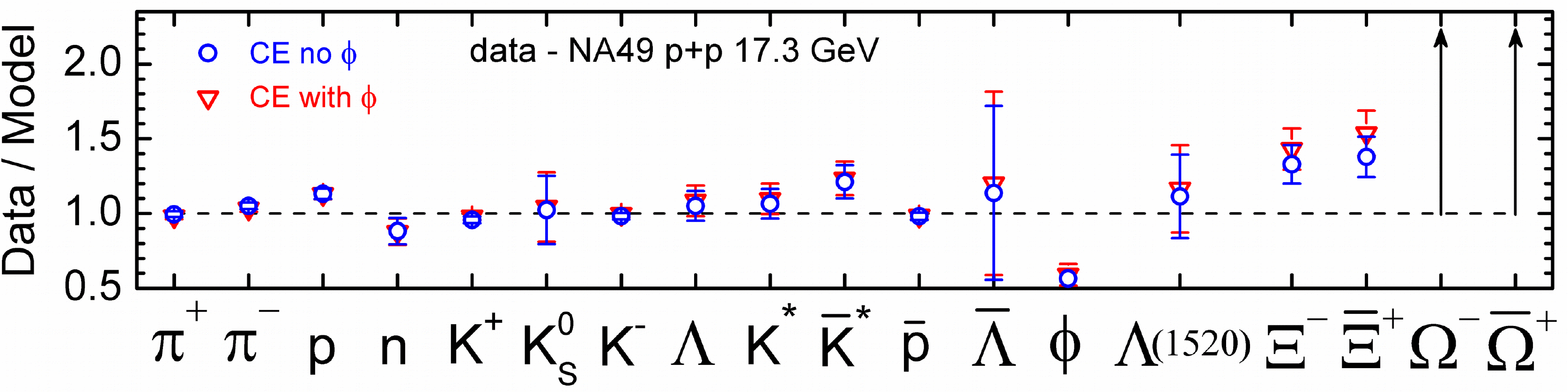}
\caption{The data over model ratios for the CE HRG model fit to the NA49 data on the $4\pi$ hadron multiplicities in inelastic p+p interactions at $\sqrt{s_{NN}}=17.3~$GeV. For $\Omega^-$ the data to model ratios are $5.9\pm 2.9$ in the `no $\phi$' fit and $6.7\pm 3.3$ `with $\phi$'. The corresponding values for $\bar{\Omega}^+$ are, respectively, $6.0 \pm 3.4$ and $7.2\pm 4.0$.
} \label{fig-NA49}
\end{figure}

We have also performed the fit of the p+p data of the NA49 Collaboration.
These data are available only at the $E_{\rm lab}=158$~GeV~($\sqrt{s_{NN}}=17.3~$GeV)~\cite{NA49-p+p} collision energy, and they include a larger set of different hadron species. 
The fit results are shown in Fig.~\ref{fig-NA49}, and the corresponding model parameters are presented in Table~\ref{Table:IV}.
Adding more multiplicity data points to the fit leads to a decrease of the error bars for fitted thermal parameters, i.e., a larger number of fitted particle species gives stronger constrains on the fit results.
However, the CE HRG model fit of the measured $\phi$-meson yields is still quite bad. Besides, an exclusion of the $\phi$ meson from the fit does not influence essentially its calculated abundance.
Large values of the data/model ratios for $\Omega^-$ and $\bar{\Omega}^+$ yields is observed: $6.7\pm 3.3$ and $7.2\pm 4.0$ correspondingly, see Table~\ref{Table:II}.
It does not fit the deviation scale of other particles, and is therefore indicated with the arrows in Fig.~\ref{fig-NA49}.
The CE HRG model results overestimate the $\phi$-meson yield and underestimate the yields of $\Omega^-$ and $\bar{\Omega}^+$. The difficulties in descriptions of the $\phi$-meson yield within the CE HRG in high energy p+p and e$^+$+e$^-$ collisions were also pointed out in Refs.~\cite{Bec,Becattini:2008,Becattini:2010}.
Our analysis shows that these difficulties persists also for p+p collisions at low SPS energies. Note that a disagreement of the $\phi$ and $\Omega$ yields with the strangeness canonical ensemble  has been
also observed for the  peripheral p+Pb collisions at 5.02~TeV at the LHC~\cite{Sharma:2018owb}.

\begin{figure}[ht!]
\centering
\includegraphics[width=0.7\textwidth]{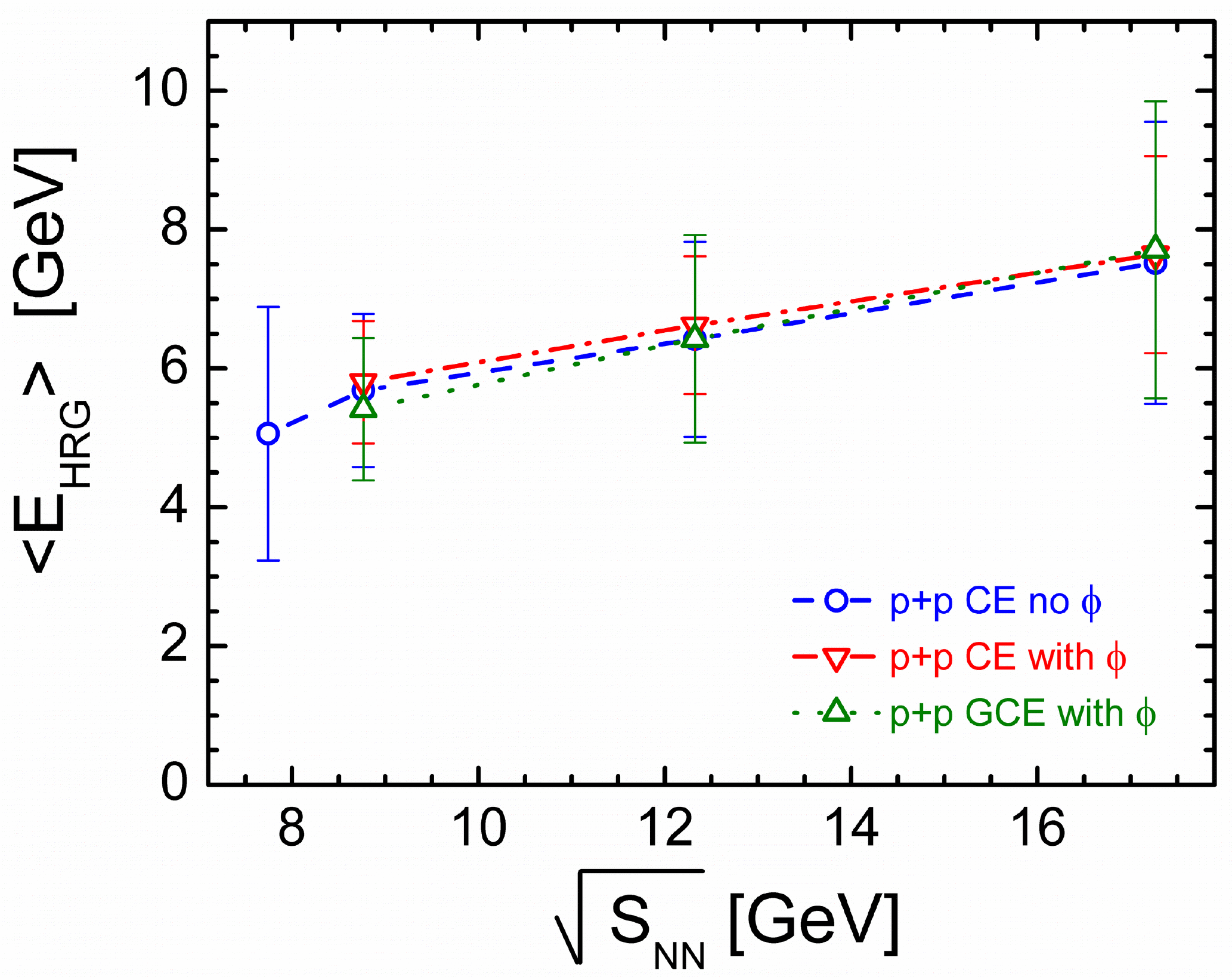}
\caption{The average energy of the statistical system calculated
with the parameters of the CE and GCE HRG models presented in Table~\ref{Table:III}.
} \label{fig-Etot}
\end{figure}

As a possible solution it was pointed out in Refs.~\cite{Becattini:2004rq,Bec} that the CE might not be adequate, because the total energy of the statistical system created in p+p collisions at the SPS energies is rather low.
Thus, the exact energy conservation may play an important role, and one should use the micro canonical ensemble (MCE)~\cite{Becattini:2004rq,Bec}.
In Fig.~\ref{fig-Etot} we present the average energy of statistical systems calculated at different collision energies with the CE parameters presented in Table~\ref{Table:III}.
The total energy of statistical system is indeed rather small, $5-8$~GeV, and approximately equals to
a half of the total available energy $\sqrt{s_{NN}}$ (the rest of the energy can be attributed to the particle motion along the collision axis).
However, the MCE calculations at the total energy of several GeV show an opposite behavior: The results of Ref.~\cite{Begun:2005qd} demonstrate the MCE {\it enhancement} of heavy ($m\gg T$) neutral particles above the threshold, i.e., at $E>m$, in a comparison with the GCE and CE results at the same volume $V$ and energy $\langle E \rangle =E_{\rm mce}$.
This intuitively unexpected behavior comes from the fact that at fixed energy $E$ and conserved charges $B$, $Q$, and $S$ the MCE suppresses more and more micro states with decreasing of $E$. These micro states are however permitted in the GCE due to the charge and energy fluctuations. To have the same energy density, the MCE should compensate these forbidden states and it produces more permitted ($m<E$) heavy ($m\gg T$) neutral states.
Therefore, the MCE effects would make the observed issues for the $\phi$-meson even worse for a {\it single} MCE cluster.

As we already pointed in the Introduction, an alternative scenario, namely that not one, but two or even more small statistical micro-subclusters are formed in inelastic p+p interactions, with masses and charges fluctuating from event to event, can not be excluded. Each of such micro-subclusters in any given event may then be describable by the MCE with exact conservation of the subcluster's energy and conserved charges.
These MCE subclusters with parameters {\it fluctuating} event-by-event may then be represented as - and look like - a single statistical system, if special conditions of CE or GCE are fulfilled.
\\

2. Indication for a presence of {\it large non-statistical (dynamical) fluctuations} in inelastic p+p interactions.

The CE, and especially MCE, faces another well known issue in considering event-by-event particle number fluctuations.
The multiplicity distributions in p+p at high energy are much broader than predicted by statistical models.
For example, the measured scaled variance for the fluctuations of negatively charged  particles, $\omega[N_-]\equiv (\langle N_-^2\rangle- \langle N_-\rangle)^2/\langle N_-\rangle$, in inelastic p+p reactions equals approximately 1.4 at $\sqrt{s_{NN}}\cong 17.3$~GeV~\cite{omega-}. The CE and MCE statistical fluctuations are essentially smaller: $\omega_{\rm ce}[N_-]\cong 0.8$ and $\omega_{\rm mce}[N_-]\cong 0.4$ \cite{ce-mce}. Thus, the p+p data do not show the expected suppression of the particle number fluctuations due to either exact charge or both charge and energy conservation. Even a presence of the GCE statistical fluctuations of both the energy and conserved charges appears to be not enough and leads to the value of $\omega_{\rm gce}[N_-]\cong 1.1$ \cite{ce-mce} which is still smaller than the p+p data. An inclusion of the excluded volume type interactions would also be expected to suppress the particle number fluctuations~\cite{Gorenstein:2007ep}. Therefore, it seems that one needs some large non-statistical fluctuations of energy $E$ and/or volume $V$ to explain the measured particle number fluctuations in p+p reactions. Note that a role of the non-statistical fluctuations should become much stronger at the RHIC and LHC energies: the p+p data show the behavior $\omega[N_-]\propto \langle N_-\rangle \gg 1$, whereas $\omega_{\rm gce}$ is only slightly larger than 1, and both $\omega_{\rm ce}$ and $\omega_{\rm mce}$ are always smaller than 1.
The MCE with strong volume fluctuations was suggested in Ref.~\cite{Begun:2008fm,Begun:2008gg}.  %
As one can see from Fig.~\ref{fig-Etot}, the value of $\langle E\rangle$ equals approximately a half of the total available energy $\sqrt{s_{NN}}$.
It leaves enough room for strong energy fluctuations of the
statistical system formed in p+p collisions at the SPS energies.
\\

3.  {\it No indications for an existence of a single statistical system} in
inelastic p+p interactions.

The fit of the hadron yields within the GCE assumes that both the system energy and conserved charges fluctuate from event to event. We use the GCE to fit the p+p data of NA61/SHINE. These fits are shown in Figs.~\ref{fig-T}, and \ref{fig-Etot}. Quite unexpectedly we observe that the quality of the GCE fits appears to be better than of the CE fits. This is clearly seen in Fig.~\ref{fig-T}~(d). Most striking advantages of the GCE are observed at large SPS energies for the fit `with $\phi$'. Better description of experimental yields is achieved within the GCE due to smaller values of the parameter $\gamma_S$. These smaller GCE values of $\gamma_S$ compensate the absence of the CE suppression effects for strange and multi-strange hadrons in the GCE.

\begin{figure}[ht!]
\centering
\includegraphics[width=1.0\textwidth]{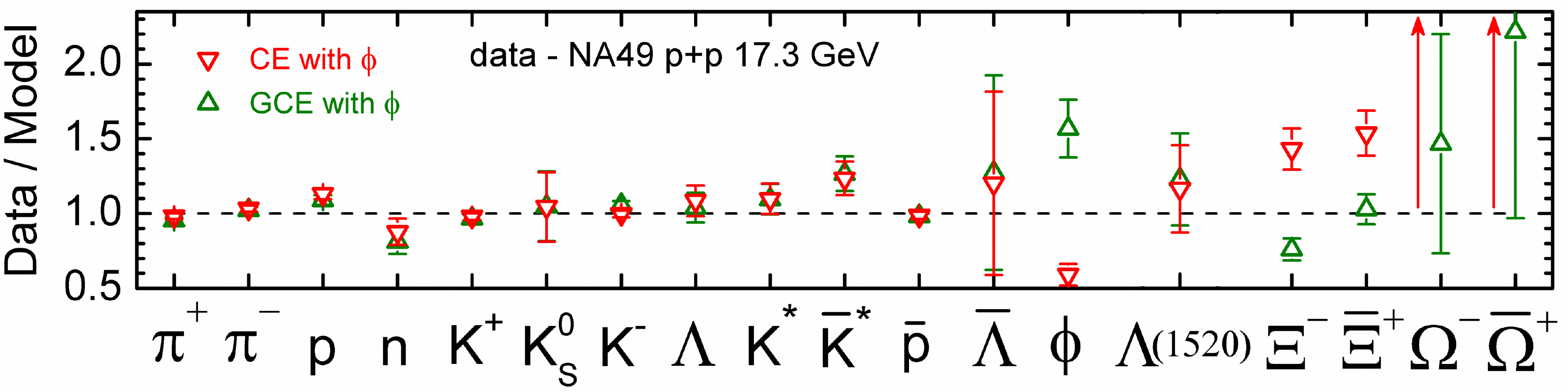} 
\caption{
The data over model ratios for the CE HRG and GCE HRG model fits to the NA49 data on the $4\pi$ hadron multiplicities in inelastic p+p interactions at $\sqrt{s_{NN}}=17.3~$GeV. For the CE the results are the same as in Fig.~\ref{fig-NA49}. The parameters of the CE and GCE fits are presented in Table~\ref{Table:IV}.} \label{fig-CEGCE}
\end{figure}

Much smaller $\chi^2/N_{\rm dof}$ in the GCE in comparison with the CE means that we have no indications in favor of the existence of a single statistical system with exactly conserved charges from one event to another. We compare the fit of the NA49 data for all particle species  within the CE presented in Fig.~\ref{fig-NA49} and the  GCE fit. The results  are shown in Fig.~\ref{fig-CEGCE}. The arrows indicate the deviation of the CE fit results. The parameters of the CE and GCE fits are presented in Table~\ref{Table:IV}.
We conclude that the GCE HRG fit of the total hadron yields in inelastic p+p  collisions at $\sqrt{s_{\rm NN}}=17.3$~GeV has an advantage in a comparison with CE HRG fit.

Possible explanation of this unexpected observation  may be found in the presence of leading baryon(s), like $p$, $n$, $\Lambda$, etc., which leads to the event-by-event fluctuations of the conserved
charge  of the remaining `statistical system'.
More generally, it is quite possible also that not one, but two or more statistical clusters (fireballs) are formed in inelastic p+p interactions, with their masses and charges fluctuating from one event to another.
Each of these fireballs in a given event can be described by the MCE with exact conservation of its energy and conserved charges. These MCE clusters with fluctuating parameters can be then represented as a single statistical system. However, the statistical ensemble which corresponds to this statistical system would not correspond to the standard MCE, CE, and, GCE, and it will depend crucially on the p+p reaction dynamics. It nevertheless looks interesting that the GCE seems to be a better approximation than the CE for this statistical system with `dynamically fluctuated' clusters.

\section{Summary}\label{sum}

The CE HRG model is used to describe the new data of the  NA61/SHINE collaboration on hadron multiplicities in inelastic p+p interactions at the CERN SPS.
The chemical freeze-out parameters, $T$, $\gamma_S$, and $R$, in p+p collisions are found and compared to those in A+A collisions.
Rather similar values of the temperature parameter in both cases are found, in agreement with our earlier studies.

The analysis, however, reveals at least three intricacies with the description of the data within the CE, GCE, or MCE formulation of the statistical model applied to the system as a whole:
\begin{enumerate}
 \item the $\phi$ meson multiplicities in p+p inelastic reactions can not be well fitted within the CE formulation of the HRG model at the SPS energies;
 \item the experimental event-by-event particle number fluctuations appear to be much larger than their values calculated within the CE and MCE of the HRG model;
\item an advantage of the GCE in comparison with the CE fits gives no indications of an existence of the {\it single} statistical system with exactly conserved charges in each event.
\end{enumerate}
The MCE applied to a {\it single} fireball does not help, but makes all these intricacies even larger. One possible way to preserve the statistical model approach to the hadron multiplicity data in p+p reactions is to introduce significant {\it non-statistical} (dynamical) {\it fluctuations} of the statistical model parameters. Two or more statistical fireballs with fluctuating energies, volumes, and the values of conserved charges should be then considered.

Presence of these non-statistical fluctuations can be tested experimentally, analyzing only the `most central' p+p collision events.
For the most central p+p events the non-statistical fluctuations are expected to become essentially smaller.
One will be able then to study whether the CE (or even the MCE) can be adequate to describe the data for both the mean hadron multiplicities and the multiplicity fluctuations in the most central p+p collisions at the SPS energies.
A corresponding centrality selection in p+p reactions can be done using the forward energy trigger similar to that used by NA61/SHINE in defining the centrality classes in A+A collisions. High statistics available for p+p collisions could make it possible to obtain the multiplicities of identified particles in different centrality bins.
This may help to understand the equilibration processes in heavy ion and elementary particle collisions at the SPS energies.

\begin{acknowledgments}
We are thankful to M. Gazdzicki, K.~Grebieszkow, A.~Motornenko, M.~S\l{}odkowski, J.~Steinheimer, and all participants of the NA61/SHINE theory meeting for fruitful discussions and comments.
The work of M.I.G. was supported by the Program of Fundamental Research of the National Academy of Sciences of Ukraine. H.St. appreciates the support from J.M.~Eisenberg Laureatus Professorship of the Fachbereich Physik der Goethe Universitaet and of the Walter Greiner Gesellschaft zur Foerderung der physikalischen Grundlagenforschung Frankfurt.

\end{acknowledgments}

%
%
\begin{table}[ht!]
\centering
\renewcommand{\arraystretch}{0.7}
\begin{tabular}{|c|c|c|c|c|c|c|}
\hline
\multicolumn{7}{|c|}{NA61/SHINE~~CE}\\
\hline
&\multicolumn{3}{|c|}{$\sqrt{s_{NN}}=8.8$ GeV} & \multicolumn{3}{|c|}{$\sqrt{s_{NN}}=12.3$ GeV} \\
\hline
 & Data & Fit no $\phi$ & with $\phi$ & Data & Fit no $\phi$ & with $\phi$\\
\hline
\hline
$\pi^+$ & $2.39 \pm 0.16$ & $2.50$ & $2.77$  & $2.67 \pm 0.14$ & $2.71$ & $3.07$ \\
$\pi^-$ & $1.71 \pm 0.17$ & $1.70$ & $1.94$  & $2.03 \pm 0.17$ & $1.92$ & $2.24$ \\
$K^+$ & $0.170 \pm 0.025$ & $0.172$ & $0.116$ & $0.201 \pm 0.014$ & $0.193$ & $0.139$ \\
$K^-$ & $0.0840 \pm 0.0067$ & $0.0782$ & $0.0568$  & $0.0950 \pm 0.0064$ & $0.098$ & $0.076$ \\
$p$ & $1.095 \pm 0.090$ & $1.044$ & $1.060$  & $1.093 \pm 0.070$ & $1.035$ & $1.050$ \\
$\bar{p}$ & $0.00590 \pm 0.00072$ & $0.00598$ & $0.00537$  & $0.01830 \pm 0.00180$ & $0.01820$ & $0.01710$ \\
$\Lambda$ & $0.082 \pm 0.010$ & $0.092$ & $0.060$  & & $0.096$ & $0.064$ \\
$\bar{\Lambda}$ & & $17.7\cdot 10^{-4}$ & $9.06 \cdot 10^{-4}$  & & $0.00533$ & $0.00286$ \\
$\Sigma^+$ & & $0.0288$ & $0.0181$  & & $0.0294$ & $0.0190$ \\
$\bar{\Sigma}^+$ & & $3.85 \cdot 10^{-4}$ & $2.00 \cdot 10^{-4}$ & & $11.5\cdot 10^{-4}$ & $6.31 \cdot 10^{-4}$ \\
$\Sigma^-$ & & $0.0172$ & $0.0114$ & & $0.0180$ & $0.0124$ \\
$\bar{\Sigma}^-$ & & $5.28 \cdot 10^{-4}$ & $2.65 \cdot 10^{-4}$ & & $15.7\cdot 10^{-4}$ & $8.26 \cdot 10^{-4}$ \\
$\Xi^0$ & & $17.2\cdot 10^{-4}$ & $6.57 \cdot 10^{-4}$ & & $19.1\cdot 10^{-4}$ & $7.71 \cdot 10^{-4}$ \\
$\bar{\Xi}^0$ & & $8.31 \cdot 10^{-5}$ & $2.32 \cdot 10^{-5}$ & & $25.2 \cdot 10^{-5}$ & $7.43 \cdot 10^{-5}$ \\
$\Xi^-$ & & $14.7\cdot 10^{-4}$ & $5.66 \cdot 10^{-4}$ & & $16.6\cdot 10^{-4}$ & $6.77 \cdot 10^{-4}$ \\
$\bar{\Xi}^-$ & & $9.10 \cdot 10^{-5}$ & $2.52 \cdot 10^{-5}$ & & $27.4 \cdot 10^{-5}$ & $7.98 \cdot 10^{-5}$ \\
$\Omega$ & & $17.8 \cdot 10^{-6}$ & $3.86 \cdot 10^{-6}$ & & $24.4 \cdot 10^{-6}$ & $5.62 \cdot 10^{-6}$ \\
$\bar{\Omega}$ & & $~2.70 \cdot 10^{-6}~$ & $~0.387 \cdot 10^{-6}~$ & & $~9.23 \cdot 10^{-6}~$ & $~1.38 \cdot 10^{-6}~$ \\
$\pi^0$ & & $2.41$ & $2.68$ & & $2.64$ & $3.00$ \\
$K^0_S$ & & $0.115$ & $0.0806$ & & $0.137$ & $0.102$ \\
$\eta$ & & $0.170$ & $0.149$ & & $0.205$ & $0.193$ \\
$\omega$ & & $0.189$ & $0.199$ & & $0.257$ & $0.283$ \\
$K^{*+}$ & & $0.0477$ & $0.0293$& & $0.0635$ & $0.0414$ \\
$K^{*-}$ & & $0.0182$ & $0.0126$ & & $0.0261$ & $0.0195$ \\
$K^{*0}$ & & $0.0377$ & $0.0237$ & & $0.0512$ & $0.0343$ \\
$\bar{K^{*0}}$ & & $0.0214$ & $0.0145$ & & $0.0303$ & $0.0222$ \\
$\rho^+$ & & $0.285$ & $0.297$ & & $0.363$ & $0.393$ \\
$\rho^-$ & & $0.174$ & $0.192$ & & $0.224$ & $0.259$ \\
$\rho^0$ & & $0.255$ & $0.269$ & & $0.325$ & $0.357$ \\
$\eta'$ & & $0.0153$ & $0.0130$  & & $0.0219$ & $0.0195$ \\
$\phi$ &$(5.34\pm 0.65)\cdot 10^{-3}$ & $14.2\cdot 10^{-3}$ & $7.91\cdot 10^{-3}$ &$(8.43\pm 0.57)\cdot 10^{-3}$& $17.8\cdot 10^{-3}$ & $10.4\cdot 10^{-3}$ \\
$\Lambda(1520)$ & & $0.00644$ & $0.00374$  & & $0.00829$ & $0.00490$  \\
\hline
\end{tabular}\caption{The CE HRG fit to the NA61/SHINE p+p data at $\sqrt{s_{NN}}=8.8$ and $12.3$ GeV.}
\label{Table:I}
\end{table}
%
%
\begin{table}[h!]
\centering
\renewcommand{\arraystretch}{0.7}
\begin{tabular}{|c|c|c|c|c|c|c|}
\hline
\multicolumn{4}{|c|}{NA61/SHINE~~CE} & \multicolumn{3}{|c|}{NA49~~CE} \\
\hline
&\multicolumn{3}{|c|}{$\sqrt{s_{NN}}=17.3$ GeV} & \multicolumn{3}{|c|}{$\sqrt{s_{NN}}=17.3$ GeV} \\
\hline
 & Data & Fit no $\phi$ & with $\phi$ & Data & Fit no $\phi$ & with $\phi$\\
\hline
$\pi^+$ & $3.11 \pm 0.26$ & $3.12$ & $3.49$ & $3.02 \pm 0.06$ & $3.03$ & $3.08$ \\
$\pi^-$ & $2.40 \pm 0.18$ & $2.33$ & $2.66$ & $2.360 \pm 0.047$ & $2.247$ & $2.285$ \\
$K^+$ & $0.234 \pm 0.022$ & $0.237$ & $0.166$ & $0.227 \pm 0.005$ & $0.237$ & $0.232$ \\
$K^-$ & $0.132 \pm 0.014$ & $0.134$ & $0.100$ & $0.130 \pm 0.003$ & $0.132$ & $0.130$ \\
$p$ & $1.154 \pm 0.041$ & $1.028$ & $1.050$ & $1.162 \pm 0.035$ & $1.027$ & $1.029$ \\
$\bar{p}$ & $0.0402 \pm 0.0036$ & $0.0409$ & $0.0382$ & $0.039 \pm 0.001$ & $0.040$ & $0.040$ \\
$\Lambda$ & $0.120 \pm 0.012$ & $0.108$ & $0.070$ & $0.116 \pm 0.011$ & $0.110$ & $0.107$ \\
$\bar{\Lambda}$ & & $0.0118$ & $0.0062$ & $0.014 \pm 0.007$ & $0.012$ & $0.011$ \\
$\Sigma^+$ & & $0.0325$ & $0.0201$ & & $0.0332$ & $0.0321$ \\
$\bar{\Sigma}^+$ & & $0.00259$ & $0.00137$ & & $0.00263$ & $0.00250$ \\
$\Sigma^-$ & & $0.0211$ & $0.0137$ & & $0.0213$ & $0.0207$ \\
$\bar{\Sigma}^-$ & & $0.00343$ & $0.00175$ & & $0.00350$ & $0.00332$ \\
$\Xi^0$ & & $0.00252$ & $0.00091$ & & $0.00263$ & $0.00244$ \\
$\bar{\Xi}^0$ & & $5.80 \cdot 10^{-4}$ & $1.59 \cdot 10^{-4}$ & & $~6.21 \cdot 10^{-4}~$ & $~5.58 \cdot 10^{-4}~$ \\
$\Xi^-$ & & $~22.4\cdot 10^{-4}~$ & $~8.13 \cdot 10^{-4}~$ & $0.0031 \pm 0.0003$ & $0.0023$ & $0.0022$ \\
$\bar{\Xi}^-$ & & $6.22 \cdot 10^{-4}$ & $1.69 \cdot 10^{-4}$ & $(9.2\pm 0.9)\cdot 10^{-4}$ & $6.67\cdot 10^{-4}$ & $5.99\cdot 10^{-4}$ \\
$\Omega$ & & $40.9 \cdot 10^{-6}$ & $7.85 \cdot 10^{-6}$ & $~(26.0\pm 13.0)\cdot 10^{-5}~$ & $4.42\cdot 10^{-5}$ & $3.9\cdot 10^{-5}$ \\
$\bar{\Omega}$ & & $23.3 \cdot 10^{-6}$ & $3.11 \cdot 10^{-6}$ & $(16.0\pm 9)\cdot 10^{-5}$ & $2.65\cdot 10^{-5}$ & $2.23\cdot 10^{-5}$ \\
$\pi^0$ & & $3.09$ & $3.46$ & & $3.00$ & $3.05$ \\
$K^0_S$ & & $0.177$ & $0.128$ & $0.18 \pm 0.04$ & $0.176$ & $0.172$ \\
$\eta$ & & $0.254$ & $0.241$ & & $0.249$ & $0.250$ \\
$\omega$ & & $0.333$ & $0.369$ & & $0.322$ & $0.327$ \\
$K^{*+}$ & & $0.0831$ & $0.0532$ & & $0.0841$ & $0.0815$ \\
$K^{*-}$ & & $0.0383$ & $0.0276$ & & $0.0378$ & $0.0372$ \\
$K^{*0}$ & & $0.0691$ & $0.0451$ & $0.0741 \pm 0.0069$ & $0.0696$ & $0.0675$ \\
$\bar{K^{*0}}$ & & $0.0436$ & $0.0308$ & $0.0523 \pm 0.0048$ & $0.0432$ & $0.0424$ \\
$\rho^+$ & & $0.444$ & $0.484$ & & $0.433$ & $0.438$ \\
$\rho^-$ & & $0.287$ & $0.331$ & & $0.277$ & $0.282$ \\
$\rho^0$ & & $0.403$ & $0.445$ & & $0.392$ & $0.397$ \\
$\eta'$ & & $0.0294$ & $0.0261$ & & $0.0290$ & $0.0289$ \\
$\phi$ & $~0.0116\pm 0.0005~$ & $0.0224$ & $0.0126$ & $0.0130\pm 0.0016$ & $0.0230$ & $0.0220$ \\
$\Lambda(1520)$ & & $0.0104$ & $0.0059$ & $0.012 \pm 0.003$ & $0.0108$ & $0.0103$ \\
$n$ & & 0.762 & 0.812 & $0.665 \pm 0.067$ & $0.754$ & $0.759$ \\
\hline
\end{tabular}\caption{The CE HRG fit to the NA61/SHINE and NA49 p+p data at $\sqrt{s_{NN}}=17.3$ GeV.}
\label{Table:II}
\end{table}

\begin{table}
\centering
\renewcommand{\arraystretch}{0.7}
\begin{tabular}{|c|c|c|c|c|c|c|c|}
\hline
\multicolumn{8}{|c|}{NA61/SHINE~~CE}\\
\hline
$\sqrt{s_{NN}}$ (GeV) & $p_{lab}$ (GeV/c) & particle set& $T$ (MeV) & $R$ (fm) & $\gamma_s$ & \multicolumn{2}{|c|}{$\chi^2/N_{\rm dof}$}\\
\hline
\hline
7.74 & 31 & no $\phi$ & 149.0 $\pm$ 6.2 & 1.56 $\pm$ 0.19 & 0.588 $\pm$ 0.057 & 5.88/3 & 1.96 \\
\hline
8.76 & 40 & no $\phi$ & 142.9 $\pm$ 3.3 & 1.85 $\pm$ 0.12 & 0.607 $\pm$ 0.034 & 2.45/4 & 0.61 \\
 & & with $\phi$
 & $136.3	\pm	2.7$
 & $2.15	\pm	0.11$
 & $0.450	\pm	0.020$
 & $49.97/5$
 & 9.99 \\
\hline
12.32 & 80 & no $\phi$ & 161.9 $\pm$ 4.3 & 1.49 $\pm$ 0.11 & 0.547 $\pm$ 0.030 & 1.70/3 & 0.57 \\
 & & with $\phi$ & 151.8 $\pm$ 3.0 & 1.82 $\pm$ 0.09 & 0.402 $\pm$ 0.014 & 49.39/4 & 12.35 \\
\hline
17.27 & 158 & no $\phi$ & 174.2 $\pm$ 5.6 & 1.36 $\pm$ 0.12 & 0.518 $\pm$ 0.031 & 10.62/4 & 2.65 \\
 & & with $\phi$ & 162.6 $\pm$ 3.8 & 1.67 $\pm$ 0.10 & 0.370 $\pm$ 0.012 & 47.44/5 & 9.49 \\
\hline
\end{tabular}\caption{The extracted CE HRG parameters from the fit to the NA49 p+p data.}
\label{Table:III}
\end{table}
\begin{table}
\vspace{-10cm}
\centering
\renewcommand{\arraystretch}{0.7}
\begin{tabular}{|c|c|c|c|c|c|c|c|c|}
\hline
\multicolumn{9}{|c|}{NA49~~CE}\\
\hline
$\sqrt{s_{NN}}$ (GeV) & $p_{lab}$ (GeV/c) & particle set& \multicolumn{2}{|c|}{$T$ (MeV)} & $R$ (fm) & $\gamma_s$ & \multicolumn{2}{|c|}{$\chi^2/N_{\rm dof}$}\\
\hline
17.27 & 158 & no $\phi$ & \multicolumn{2}{|c|}{ 175.8 $\pm$ 1.5} & 1.31 $\pm$ 0.03 & 0.533 $\pm$ 0.009 & 51.64/14 & 3.69 \\
 & & with $\phi$ & \multicolumn{2}{|c|}{ 174.3 $\pm$ 1.5} & 1.35 $\pm$ 0.03 & 0.518 $\pm$ 0.008 & 87.06/15 & 5.80 \\
\hline
\multicolumn{9}{|c|}{NA49~~GCE}\\
\hline
$\sqrt{s_{NN}}$ (GeV) & $p_{lab}$ (GeV/c) & particle set& $T$ (MeV) & $\mu_B$ (MeV) & $R$ (fm) & $\gamma_s$ & \multicolumn{2}{|c|}{$\chi^2/N_{\rm dof}$}\\
\hline
17.27 & 158 & no $\phi$ & $165.2\pm 1.5$ & $231.2\pm 3.1$ & $1.47\pm0.04$ & $0.330\pm0.006$ & 47.35/13 & 3.64 \\
 & & with $\phi$ & $165.4\pm 1.5$ & $231.4\pm 3.1$ & $1.46\pm 0.04$ & $0.333\pm 0.006$ & 56.20/14 & 4.01 \\
\hline
\end{tabular}\caption{The extracted CE and GCE HRG parameters from the fit to the NA49 p+p data.}
\label{Table:IV}
\end{table}

\end{document}